\documentclass[aps,prb,twocolumn,amsmath,amssymb,showpacs,superscriptaddress]{revtex4-2}
\usepackage{graphicx}
\usepackage{xcolor}
\usepackage{lipsum}
\usepackage{hyperref}
\usepackage{cleveref}
\usepackage{xfrac}
\usepackage{gensymb}
\usepackage{amsmath}
\usepackage[normalem]{ulem}

\begin{document}
\title{Entangled-photon time- and frequency-resolved optical spectroscopy}
\author{Raúl Álvarez-Mendoza}
\affiliation{School of Physics and Astronomy, University of Glasgow, Glasgow, G12 8QQ, UK}
\author{Lorenzo Uboldi}
\email{RA-M \& LU contributed equally}
\affiliation{Department of Physics, Politecnico di Milano, P.zza L. da Vinci 32, 20133 Milano, Italy}

\author{Ashley Lyons}
\affiliation{School of Physics and Astronomy, University of Glasgow, Glasgow, G12 8QQ, UK}
\author{Richard J. Cogdell}
\affiliation{School of Molecular Biosciences, University of Glasgow, Glasgow, G12 8QQ, UK}
\author{Giulio Cerullo}
\email{giulio.cerullo@polimi.it}
\affiliation{Department of Physics, Politecnico di Milano, P.zza L. da Vinci 32, 20133 Milano, Italy}
\affiliation{CNR-Institute for Photonics and Nanotechnologies (CNR-IFN), Milan 20133, Italy}
\author{Daniele Faccio}
\email{daniele.faccio@glasgow.ac.uk}
\affiliation{School of Physics and Astronomy, University of Glasgow, Glasgow, G12 8QQ, UK}

\begin{abstract}
Classical time-resolved optical spectroscopy experiments are performed using sequences of ultrashort light pulses, with photon fluxes incident on the sample which are many orders of magnitude higher than real-world conditions corresponding to sunlight illumination. Here we overcome this paradigm by exploiting quantum correlations to perform time-resolved spectroscopy with entangled photons. Starting from spontaneous parametric down-conversion driven by a continuous-wave laser, we exploit the temporal entanglement between randomly generated signal/idler pairs to obtain temporal resolution, and their spectral entanglement to select the excitation frequency. We also  add spectral resolution in detection, using a Fourier transform approach which employs a common-path interferometer. We demonstrate the potential of our entangled-photon streak camera by resolving, at the single-photon level, excitation energy transfer  cascades from LH2 to LH1 in the photosynthetic membrane and disentangling the lifetimes of two dyes in a mixture. We show that time-resolved spectroscopy with quantum light can be performed without compromising measurement time, recording a fluorescence time trace in less than a minute even for  samples with low quantum yield, which can be reduced to sub-second times with acceptable signal-to-noise ratio. Our results may usher a new era in ultrafast optical spectroscopy, where experiments are performed under conditions comparable to real-world sunlight illumination. 
\end{abstract}
\maketitle
\section*{Introduction}
Optical spectroscopy is a powerful investigation tool for the microscopic mechanisms underlying various physical, chemical and biological processes. Since the development of mode-locked laser systems, ultrafast optical spectroscopy has played a pivotal role in unveiling the early snapshots of photophysical events with picosecond to femtosecond time resolution \cite{Maiuri2020}. The studied phenomena, relevant both for a fundamental physico-chemical understanding and for technological applications, include for example, charge photogeneration in photovoltaic materials \cite{Falke2014}, rate-limiting steps in photosynthesis \cite{Scholes2011}, the mechanism of human vision \cite{Polli2010} and non-equilibrium lattice dynamics in solids \cite{Mitrano2016}. 

In standard implementations of ultrafast spectroscopy, the system under study is subjected to a series of light pulses whose frequencies and time delays are the control knobs of the experiment \cite{mukamel1995principles}. The simplest experiments are performed by exciting samples with an ultrashort light pulse and detecting their time-dependent fluorescence or absorption changes. Such experiments are described using a semiclassical formulation of light–matter interaction in which light pulses are treated as electromagnetic waves, neglecting their quantum nature. They are typically performed with comparatively high photon fluxes impinging on the samples, which are very far from real-world conditions corresponding to sunlight illuminating, e.g., a photosynthetic complex or a solar cell. For example, a 100-fs pulse with a modest energy of 1 nJ focused to a diameter of 100 $\mu$m corresponds to an intensity of \(10^{12} W/m^2\) which is nine orders of magnitude higher than the typical solar irradiance (\(\approx10^{3} W/m^2\)). Reducing the excitation intensity to more realistic values would result in vanishing nonlinear signals for classical time-resolved experiments. 

Recently, a growing body of theoretical studies \cite{Dorfman2016, Schlawin2016, Fujihashi2023, kim2024practical}  has proposed the exploitation of quantum states of light to enhance ultrafast optical spectroscopy. Most of these schemes take advantage of quantum correlations, using squeezed states of light or entangled photon pairs (EPPs) generated by spontaneous parametric down-conversion (SPDC). Quantum light offers several advantages for spectroscopy \cite{mukamel2020roadmap}, such as increasing signal-to-noise ratio, providing novel control knobs (such as energy, time and polarization entanglement of the EPPs) for designing experiments or even generating completely new signals with respect to classical light. In particular, the use of quantum light can overcome the accuracy limits in measurements with classical light, which are set by the granular (shot) noise, and exploit quantum correlations to reach the so-called Heisenberg limit \cite{Giovannetti2011, taylor2016quantum}. Examples of theoretical proposals of the use of quantum light for time-resolved spectroscopy are controlling exciton pathways in molecular aggregates by absorption of EPPs \cite{Schlawin2013}, use of EPPs to probe exciton-exciton interactions \cite{Bittner2020, li_hao}, control of azobenzene photoisomerization yield via the entanglement time of EPPs \cite{Gu2021} and fluorescence-detected two-dimensional spectroscopy using EPPs to increase spectral resolution and reveal cross-peaks \cite{Raymer2013}.

Despite the plethora of theoretical proposals, experimental demonstrations of nonlinear and time-resolved spectroscopy with quantum light are, to date, very limited. Notable examples are the linear scaling with the light intensity of the two-photon absorption rate for time-frequency EPPs \cite{Varnavski2020}, the use of quantum correlations to enhance nonlinear optical microscopy/spectroscopy, pushing the sensitivity of stimulated Raman \cite{Casacio2021} or Brillouin \cite{Li2022, li2024harnessing} scattering or time-domain THz spectroscopy \cite{adamou2025quantum} below the shot-noise limit, and the use of squeezed states of light to enhance sensitivity and generate extra nonlinear signals in four-wave-mixing of rubidium vapors \cite{Dorfman2021}. 

Recently, a pioneering study by Li \textit{et al.} \cite{Li2023} used time-resolved photon-counting quantum light spectroscopy to study excitation energy transfer (EET) processes in the light-harvesting 2 (LH2) complex of the purple photosynthetic bacterium \textit{Rhodobacter sphaeroides}, containing two rings of bacteriochlorophylls (BChls), B800 and B850.  In that study, an EPP at 808 nm was generated by SPDC from a pulsed ultrashort laser. One of the two photons served as a heralding photon and was detected by a single-photon avalanche diode (SPAD) while the other was used to excite the B800 BChls in the LH2. Following EET to the B850 BChls, which occurs on the picosecond timescale \cite{ma1997energy}, the fluorescence photon emitted by the LH2 was detected by a second SPAD and its delay with respect to the heralding photon was measured by an event timer, thus reconstructing the fluorescence lifetime by time-correlated single photon counting (TCSPC). By correlating the heralding and the detected photons, this experiment demonstrated that photosynthetic EET from B800 to B850 proceeds at the single-photon level.

In this work we exploit entanglement to demonstrate frequency and sub-200-ps time-resolved fluorescence following single-photon excitation from a continuous wave (CW) laser. We extend the experiment by Li \textit{et al.} in two important ways. First, we generate the EPP starting from a CW laser and exploit the temporal entanglement between the photons generated by the SPDC process to achieve $\sim$200-ps temporal resolution and their spectral entanglement to achieve resolution in excitation frequency. We then resolve the spectrum of the fluorescence photons using a time-domain Fourier transform approach, employing an ultra-stable birefringent interferometer \cite{Brida2012}. We thus exploit entanglement to measure time-resolved fluorescence spectra following single-photon excitation from a CW laser and, by combining temporal and spectral resolution, achieve an “entangled photon streak camera”. We demonstrate the potential of this spectroscopic tool in two experiments: we resolve, at the single-photon level, EET cascades within a photosynthetic membrane and disentangle the lifetimes of two dyes in a mixture. Critically, we highlight the broad impact and applicability of our approach beyond a proof-of-principle demonstration, by recording TCSPC time traces of light-harvesting complexes with acquisition times down to sub-second.

%

%
\section*{Experimental setup}
\begin{figure}
    \centering
    \includegraphics[width = \linewidth]{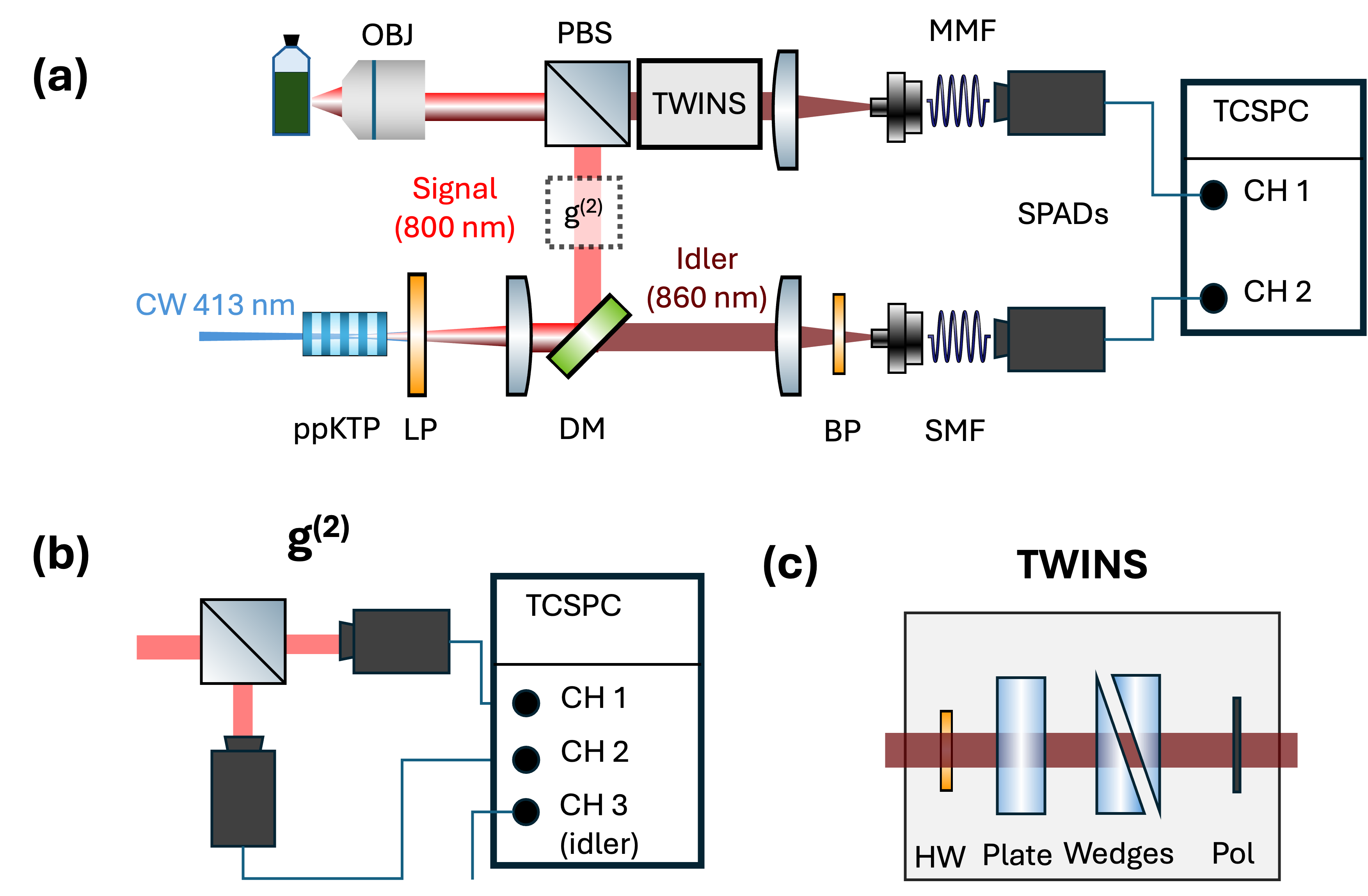}
    \caption{\textbf{Time- and frequency resolved spectroscopy with EPPs.} (a) Experimental setup. LP: long-pass filter; DM: dichroic mirror; PBS: polarizing beam splitter; BP: bandpass filter; OBJ: microscope objective; SMF: single-mode fiber; MMF: multi-mode fiber; SPAD: single-photon avalanche diode; TCSPC: time-correlated single photon counting. (b) Setup for \(g^{(2)}\) measurement. (c) Scheme of the TWINS interferometer. HW: half-waveplate; Plate: $\alpha$-BBO plate; Wedges: movable $\alpha$-BBO wedges; Pol: polarizer.}
    \label{fig:setup}
\end{figure}

Figure \ref{fig:setup}(a) shows the experimental setup. It starts with a CW single-frequency blue laser ($\lambda$ = 413 nm) focused in a 30-mm-long periodically poled KTiOPO\(_4\) (ppKTP) crystal designed for type 0 quasi phase matching, with poling period $\Lambda$ = 3.675 $\mu$m. Typical laser power is 0.25 mW and the beam is collimated to a diameter of 400 $\mu$m, corresponding to a confocal parameter of 600 mm. At 56$\degree$, ppKTP generates EPPs at 800 and 860 nm, which are separated by a dichroic mirror (DM in Fig. 1(a)). The 860-nm photon, which acts as heralding photon, is selected by an interferential bandpass filter (BP in Fig. \ref{fig:setup}(a)) and is coupled via a single-mode fiber (SMF in Fig. \ref{fig:setup}(a)) to a SPAD. The 800-nm photon is focused via a polarizing beam splitter and a 0.65-numerical aperture microscope objective to a cuvette containing the molecular sample to be investigated. The sample fluorescence is collected in the back-scattering direction via the same microscope objective and then coupled via a graded index multi-mode fiber (MMF in Fig. 1(a)) to a second SPAD. The two SPAD outputs are sent to a TCSPC unit which monitors the delay of the fluorescence photon with respect to the heralding photon and, by building a histogram of these delays, measures the fluorescence lifetime. Note that, differently from classical TCSPC systems which use a pulsed laser for photoexcitation \cite{lakowicz2006principles}, here the excitation laser is CW and the time resolution is provided by the temporal entanglement between the heralding and the illuminating photon.

To add spectral resolution to the system we use Fourier transform detection, by placing in front of the second SPAD a birefringent interferometer, the Translating-Wedge-based Identical pulses eNcoding System (TWINS, Fig. \ref{fig:setup}(c), see Methods for a detailed description) \cite{Brida2012}, which creates two delayed replicas of the incident optical waveform with interferometric delay stability. By recording a series of TCSPC traces for different replicas delays and performing a Fourier transform, one obtains a time-resolved fluorescence spectrum following single-photon excitation. Importantly, the use of TWINS implies that the fluorescence photon can be efficiently collected on a simple bucket SPAD detector rather than with a SPAD array (or a scanning SPAD detector) that would instead be required with a grating spectrometer, leading to higher losses, lower signal-to-noise ratio (the photons are spread across multiple SPADs) and more complicated detection hardware.

\section*{Results}
\begin{figure*}
    \centering
    \includegraphics[width=1\textwidth]{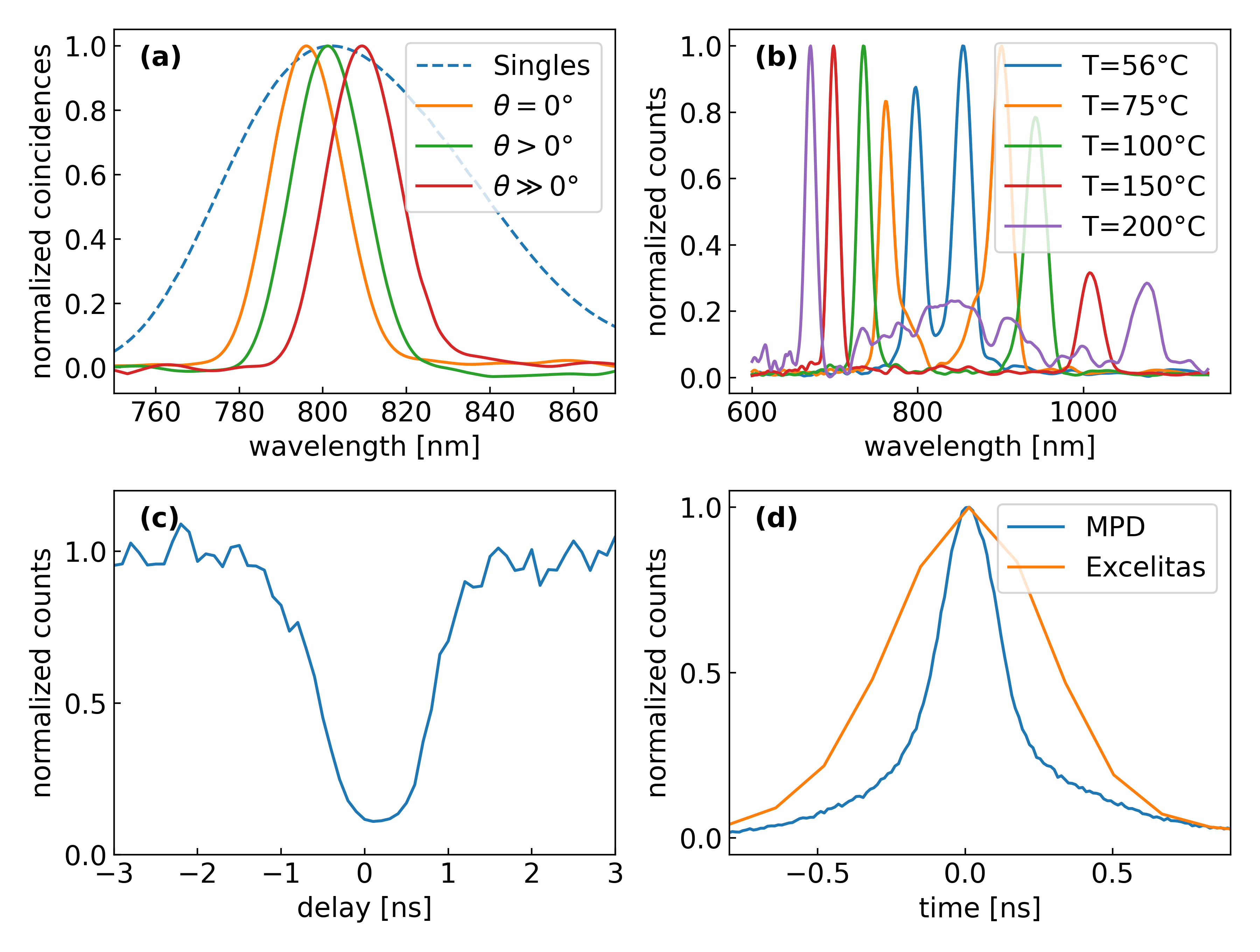}
    \caption{\textbf{Generation of frequency-tunable entangled photon pairs.} (a) solid line: spectrum of the singles generated by the ppKTP crystal at 56°C, measured with a TWINS interferometer in front of the SPAD; dashed lines: spectra of the coincidences with an interference filter on the second detector tilted by different angles; (b) SPDC spectra generated by the ppKTP crystal for different temperatures, showing tunability from 670 nm to 1100 nm; (c) \(g^{(2)}\) measurement of the signal pulses performed with an HBT interferometer; (d) IRF of the system measured detecting coincidences between signal and idler pulses, with two SPADs from MPD and Excelitas. The FWHM of the response is 260 ps (MPD) and 600 ps (Excelitas) respectively.}
    \label{fig:source}
\end{figure*}

We first characterize the source of EPPs by placing the second SPAD on the signal beam path. Figure \ref{fig:source}(a) shows the spectrum of the signal detected with the TWINS for a temperature t = 56$\degree$ of the ppKTP, which is broadband because the SPDC process is close to the degenerate condition, with group velocity matching between signal and idler \cite{manzoni2016design}. We can nonlocally control the spectrum of the excitation photons that lead to detected fluorescent photons by placing on the idler an interference filter with 10-nm bandwidth at 860 nm.  The corresponding signal photon spectrum then peaks at 800 nm with 10 nm bandwidth, as shown by the orange line in Fig. \ref{fig:source}(a), and can be tuned by $\sim$20 nm by tilting the herald photon interference filter. Under these conditions, the number of coincidences is 2$\times10^5$/second.

The phase matching curves for signal and idler can be tuned out of the degeneracy condition by heating the ppKTP crystal, obtaining complete tunability of the signal from 670 nm to 1100 nm (measurements shown in Fig. \ref{fig:source}(b)). Even broader tunability towards the green/blue could in principle be obtained by using a crystal with multiple poling periods. One should note that we demonstrate broad tunability in the excitation wavelength by using a very simple system, which consists of a CW laser and a periodically poled crystal. This should be compared to the considerably more complex systems, such as a mode-locked oscillator pumping an optical parametric oscillator \cite{pelouch1992ti}, used for time-resolved fluorescence spectroscopy with classical light.

To check that the light exciting the sample consists of single photons, we characterized the signal beam with a heralded Hanbury-Brown-Twiss (HBT) interferometer, consisting of a beam splitter coupled to two SPADs in transmission and reflection, respectively, as shown in Figure \ref{fig:setup} (b). Figure \ref{fig:source}(c) shows the normalized second order correlation function \(g^{(2)}\) as a function of the delay between the detector readings, conditioned to the arrival of the idler photon; it displays a clear dip at zero delay indicating single-photon emission, which becomes deeper approaching zero upon reducing the pump power to the ppKTP crystal. 

Finally, we determine the time resolution and the instrumental response function (IRF) of the experiment by recording the coincidences as a function of time. Since the entanglement time of the EPPs is much shorter than the response time of the SPADs, we can consider that  the IRF is exclusively determined by the SPADs. For this experiment we used two types of SPAD: one with smaller area (50 $\mu$m\(^2\)) but faster response time (MPD PD-050-CTD-FC) and one with larger area (180 $\mu$m\(^2\)) and higher quantum efficiency but slower response time (Excelitas SPCM-800-14-FC). While the faster SPAD is always used to detect the heralding photon, which can be tightly focused due to its high beam quality, a trade-off between signal strength and response time occurs for the signal photon which, due to the incoherent nature of the fluorescence, is more challenging to refocus. Figure \ref{fig:source}(d) compares two IRFs measured using the fast and the slow detector on the signal photon. Overall, upon deconvolution of the IRF, the temporal resolution of our measurements ranges between 100 and 200 ps. 

\begin{figure*}
    \centering
    \includegraphics[width=1\textwidth]{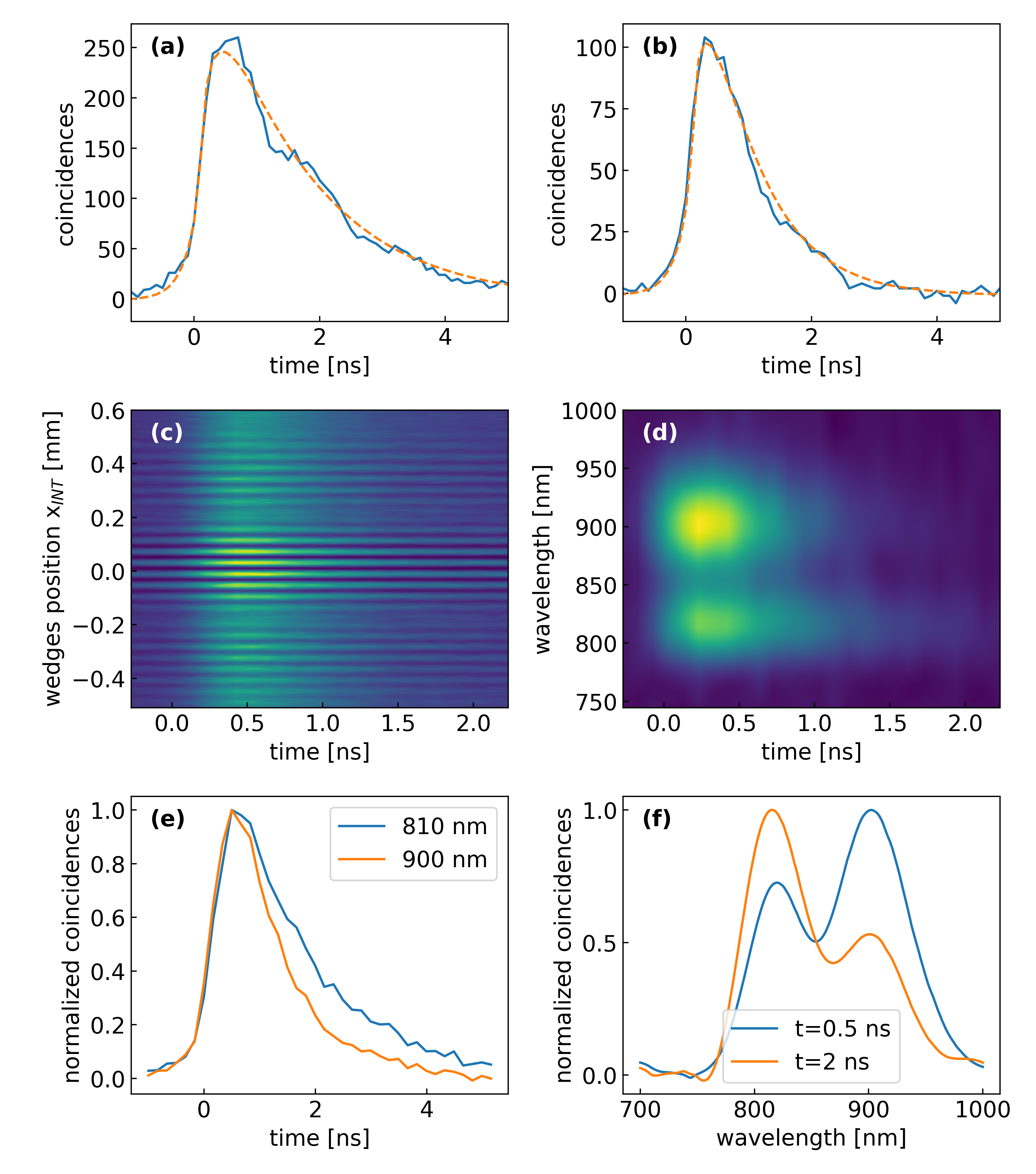}
    \caption{\textbf{Time- and frequency-resolved fluorescence spectroscopy at the single-photon level}. (a) wavelength-integrated fluorescence decay from 800CW infrared dye in DMSO following single-photon excitation at 800 nm, and mono-exponential fit with $\tau$ = 1.51 ± 0.01 ns (dashed line); (b) wavelength-integrated fluorescence decay from IR143 infrared dye following single-photon excitation at 800 nm, and mono-exponential fit with $\tau$ = 0.79 ± 0.01 ns (dashed line); (c) 2D fluorescence map as a function of emission time and wedge position x$_{INT}$ of the TWINS interferometer for a mixture of 800CW and IR143 dyes in DMSO solvent, following single-photon excitation at 800 nm. (d) time-dependent fluorescence spectra obtained by a Fourier transform of map (c) with respect to x$_{INT}$; (e) fluorescence decay dynamics at selected wavelengths; (f) fluorescence spectra at selected delays.}
    \label{fig:two_dyes}
\end{figure*}

We then send the signal photon to the sample and detect the fluorescence using the MPD SPAD. Figure \ref{fig:two_dyes}(a) shows a wavelength-integrated time-resolved fluorescence trace recorded for the infrared dye 800 CW in dimethyl sulfoxide (DMSO) solvent. One observes a mono-exponential decay with a time constant $\tau$ = 1.51$\pm$0.01 ns, in excellent agreement with classical TCSPC measurements obtained upon photoexcitation with a $\sim$100-fs pulse at 800 nm. Similarly, Figure \ref{fig:two_dyes}(b) shows the time-resolved fluorescence for the dye IR143, which displays a significantly faster decay with time constant $\tau$ = 0.79$\pm$0.01 ns. 

The acquisition time for the traces reported in Fig. \ref{fig:two_dyes}(a) and \ref{fig:two_dyes}(b) is 30 seconds. One should note that TCSPC using EPPs generated by a CW laser was recently reported \cite{harper2023entangled, eshun2023fluorescence, gabler2025benchmarking} however, with significantly lower temporal resolution and longer measurement time. Our results demonstrate that TCSPC using EPPs can be performed with similar data quality, in terms of temporal resolution and signal-to-noise ratio, to the classical measurement and without significantly compromising the measurement time, while dramatically decreasing the excitation fluence.

To demonstrate the power of our approach to characterize spectro-temporal dynamics, we measure a mixture of the two dyes 800CW and IR143, dissolved in the solvent DMSO. Both dyes absorb at 800 nm but display different fluorescence spectra, peaking at 810 and 900 nm respectively, and lifetimes (see above). To obtain resolution in the detection frequency, we insert the TWINS on the detection path and record a fluorescence decay trace for each position x$_{INT}$ of the wedge generating the replicas in the interferometer, obtaining the 2D map shown in Fig. \ref{fig:two_dyes}(c). To compensate for the insertion losses of the interferometer, we used the Excelitas SPAD for this measurement, trading temporal resolution for photon collection efficiency. The total measurement time is 120 min due to the need to record multiple time traces, one for each wedge position x$_{INT}$. By performing a Fourier transform with respect to x$_{INT}$ and applying a suitable calibration procedure described in detail in Ref. \cite{perri2018time}, we obtain the time- and frequency-resolved fluorescence map reported in Fig. 3(d). This shows that our approach allows one to obtain a remarkable amount of information on the spectro-temporal dynamics of the system following single-photon excitation. 
The time- and frequency-resolved fluorescence map of the mixture shown in Figure \ref{fig:two_dyes}(d) displays two peaks, corresponding to the two dyes, with different lifetimes, see Fig. \ref{fig:two_dyes}(e). As illustrated in Fig. \ref{fig:two_dyes}(f), the shape of the fluorescence spectrum thus evolves in time, peaking at 910 nm (820 nm) at 0.5 ns (2 ns).

\begin{figure*}
    \centering
    \includegraphics[width = \textwidth]{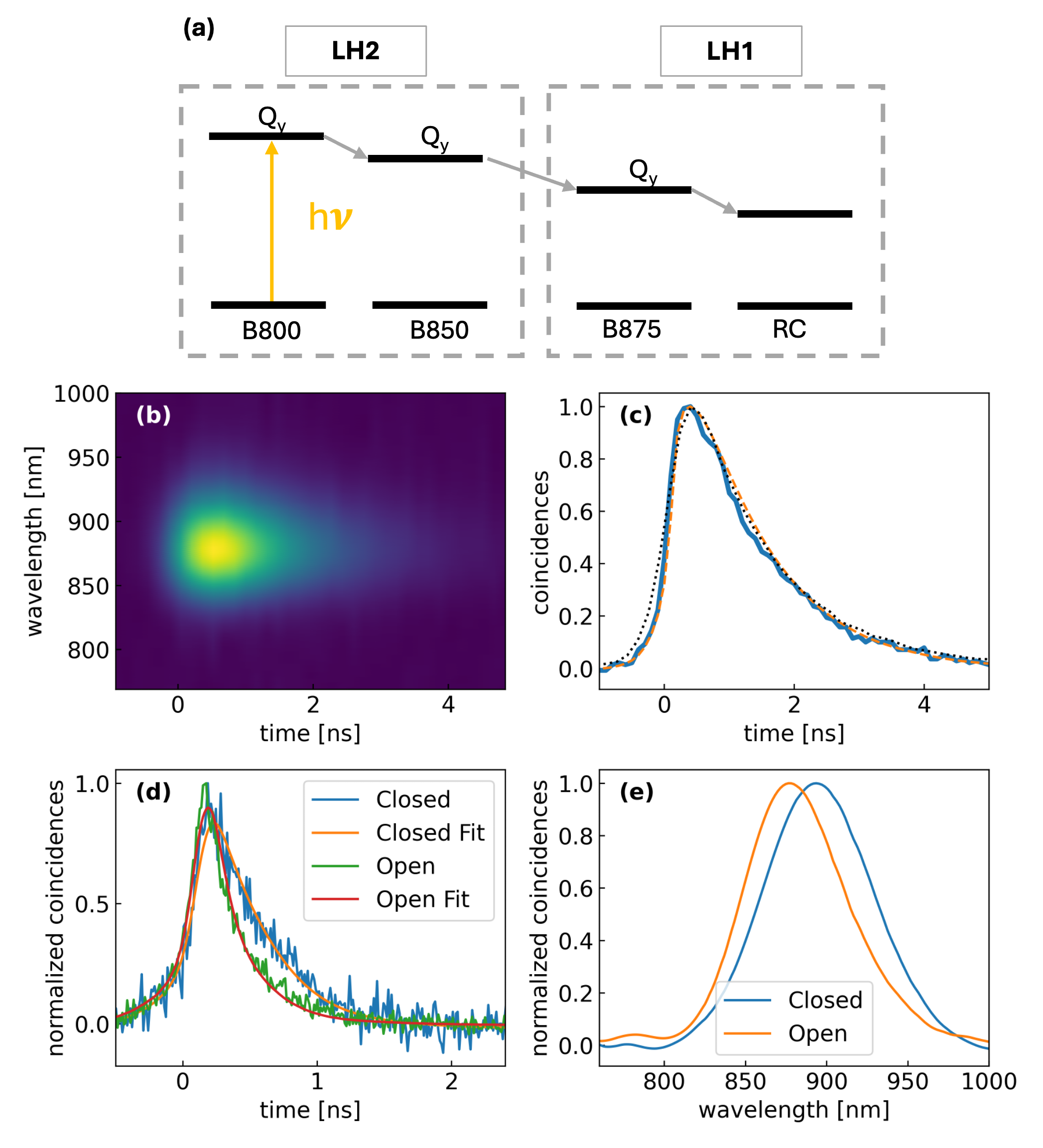}
    \caption{\textbf{Time- and frequency-resolved fluorescence spectroscopy of photosynthetic membranes at the single-photon level}. (a) energy level scheme of the BChls in the LH2 and LH1 complexes, indicating the cascade of EET processes leading to light capture in the RC and charge separation; (b) time- and frequency-resolved fluorescence spectrum of LH2 of the purple photosynthetic bacterium \textit{R. acidophilus}, following single-photon excitation of B800 BChls and EET to B850; (c) fluorescence decay at 860 nm and fit (dashed line), blue continuous line single photon excitation, black dotted line femtosecond pulse excitation; (d) wavelength-integrated time-resolved fluorescence decay of the photosynthetic membrane of \textit{Rps. sphaeroides}, following single-photon excitation at 800 nm with open (blue trace) and closed (red trace) RCs; (e) time-integrated fluorescence spectra of the photosynthetic membrane with open (blue) and closed (red) RCs.}
    \label{fig:lh}
\end{figure*}

We then characterized light-harvesting complexes of the photosynthetic purple bacterium \textit{Rhodoblastus (R.) acidophilus}. The photosynthetic membrane of a purple photosynthetic bacterium \cite{hu_pnas, Bahatyrova2004} contains peripheral antenna complexes known as LH2 complexes, which surround the core complex (LH1), containing at its center the reaction center (RC) in which charge separation takes place. As discussed in the introduction, the LH2 complex contains the B800 and B850 BChls, with absorption of the Q$_y$ transitions peaking at 800 and 850 nm, respectively. After excitation at 800 nm, EET to B850 takes place on the picosecond time scale, followed by a nanosecond fluorescence from B850 which is the terminal emitter. The LH1 complex also contains another larger ring of Bchls, called B875, with absorption red-shifted to 875 nm, thus ensuring downhill EET and energy capture from the peripheral complexes. From B875 a further EET to the RC takes place, where efficient charge separation occurs (see Fig. \ref{fig:lh}(a)).  
Figure \ref{fig:lh}(b) shows a time- and frequency-resolved fluorescence map of the LH2 from \textit{Rps. acidophilus} in buffer solution, following single-photon excitation at 800 nm. As expected, and in agreement with \cite{Li2023}, we observe a mono-exponential decay of the fluorescence from B850. Our time resolution is in fact too low to observe the $\sim$1-ps EET from B800 to B850. We do not observe a significant wavelength dependence of the decay time. Figure \ref{fig:lh}(c) shows a cut at 860 nm and reveals a time constant of 1.13$\pm$0.01 ns, in excellent agreement with classical measurements performed with a femtosecond pulsed laser (also shown in the figure).
We then moved to entire photosynthetic membranes containing both LH2 and LH1. In particular, the RCs in LH1 can be either open (i.e. functional and ready to perform charge separation) or closed (i.e. with the charge separation process blocked). Figure \ref{fig:lh}(d) shows a wavelength integrated fluorescence dynamics for a photosynthetic membrane with open RCs. In this case, the EET processes from B850 to B875 and from B875 to the RC shorten the excited state lifetime to 101$\pm$3 ps. When instead the RCs are closed (by inserting the reducing agent sodium dithionite in the buffer) the charge separation process is inhibited and the RC equilibrates with the LH1, which becomes the terminal emitter. In this case, the fluorescence lifetime increases to 248$\pm$9 ps. Figure \ref{fig:lh}(e) shows the time-integrated spectra for open and closed RCs, measured with the TWINS interferometer. When the RCs are closed, the fluorescence red-shifts, peaking at 875 nm due to the superposition of emission from B850 and B875 BChls. When the RCs are open, the charge separation process in the RCs quenches the fluorescence from B875 nm so that the residual emission, dominated by B850, is blue-shifted.

\begin{figure*}
    \centering
    \includegraphics[width=\textwidth]{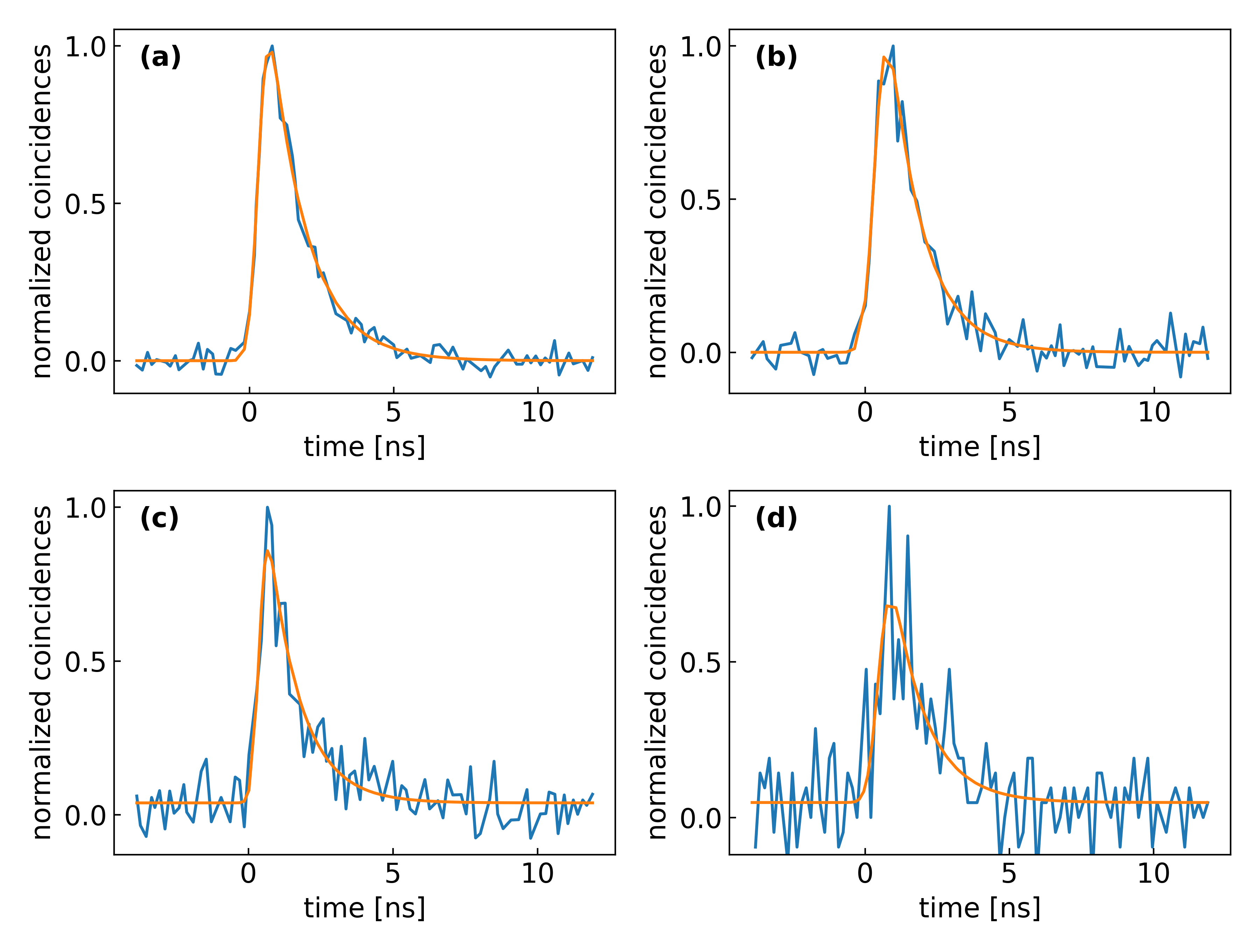}
    \caption{\textbf{LH2 lifetime estimation for various integration times.} Wavelength-integrated fluorescence decays for LH2 of the purple photosynthetic bacterium \textit{R. acidophilus}, following single-photon excitation at 800 nm, measured with different integration times: (a) 50 seconds ($\tau=1.14$ ns); (b) 10 seconds ($\tau=1.16$ ns); (c) 2 seconds ($\tau=1.11$ ns); (d) 0.6 seconds ($\tau=1.17$ ns). Orange lines are exponential fits convoluted with the IRF.}
    \label{fig:short_time}
\end{figure*}

Finally, to show the broad applicability of our approach to time-resolved spectroscopy using EPPs, we demonstrate its capability to record photoinduced dynamics without significantly compromising the measurement time with respect to classical approaches. Figure \ref{fig:short_time} shows wavelength-integrated TCSPC time traces of LH2 for various measurement times: for 50 seconds (panel a) the signal-to-noise ratio is very high, then it progressively degrades down to 2 seconds (panel c). Remarkably, even for an acquisition time of 0.6 seconds (panel d) the dynamics can be recovered, although with significantly higher noise.

\section*{Discussion}
In this work we show how quantum correlations can be used to perform time-resolved spectroscopy of ultrafast photoinduced processes under the least possible perturbative conditions. We overcome the paradigm of classical time-resolved spectroscopy, which is performed using pulsed light sources, and start from a continuous wave laser exploiting the entanglement time between randomly generated signal/idler pairs to obtain temporal resolution. Importantly, we demonstrate that time-resolved spectroscopy with quantum light can be performed without compromising measurement time, recording a fluorescence time trace in less than a minute, that can be reduced to $\sim$1 second with acceptable signal-to-noise ratio. We also show the ability to add spectral resolution to the measurement, using a Fourier transform approach which employs an interferometer on detection. While our temporal resolution is currently limited to $\sim$100$\div$200-ps by the instrumental response function of the photodetectors, it could be improved by up to one order of magnitude using suitable detectors, such as superconducting nanowire single-photon detectors. The time resolution could be further improved to ps or sub-ps levels by exploiting the Hong-Ou-Mandel (HOM) effect, whereby the fluorescence photon is made to interfere with an identical photon, also generated by SPDC \cite{lyons2023fluorescence}. Since the photon entanglement time (related to the phase matching bandwidth of the SPDC process) can be as short as hundreds of femtoseconds, one should be able to achieve sub-picosecond temporal resolution by measuring the rise and decay time of the HOM dip. In summary, we envision that the development of quantum technologies may usher a new era in ultrafast optical spectroscopy, where experiments are performed under conditions comparable to real-world sunlight illumination levels.
\section*{Methods}
\noindent \textbf{Fluorescent dye sample preparation.}
The dyes IR143 (perchlorate crystalline powder, from Exciton) and 800 CW (NHS Ester, Lycorbio) were dissolved in DMSO (purity 99\%, Sigma-Aldrich) at a concentration of $\sim1.5\cdot \text{mmol}\cdot \text{L}^{-1}$. For the measurements, $250$ $\mu$L of each dye solution was transferred into separate 1 mm path-length quartz cuvettes (CV1Q035AE, Thorlabs).

To prepare the mixed sample, $75$ $\mu$L of the 800CW solution and $150$ $\mu$L of the IR143 solution were combined in a third cuvette.

\noindent \textbf{Light harvesting complexes preparation.} 
The LH2 complexes from the purple photosynthetic bacterium \textit{Rhodoblastus acidophilus} strain 10050 (formerly known as \textit{Rhodopseudomonas acidophila}) were prepared as previously described  by Cogdell et al. \cite{cogdell1983isolation}. The purified LH2 complexes were suspended in 0.1\% v/v LDAO in 20 mM Tris HCL pH 8.0. Photosynthetic membranes from \textit{Rhodobacter sphaeroides} strain 2.4.1 were prepared from cells grown anaerobically in the light as previously described  by Cogdell et al. \cite{cogdell1975carotenoid}. The purified membranes were suspended in 20 mM MES pH 6.3, 100 mM KCl. The reaction centers in the membranes were ‘closed’ by the addition of sodium dithionite, which reduced the reaction centre’s primary electron acceptor (as described in \cite{cogdell1975carotenoid}). Photosynthetic membranes from \textit{Rb. sphaeroides} were used, rather than those from \textit{R. acidophilus}, because they are small membrane vesicles (chromatophores) that are much less light scattering than the highly light scattering membrane sheets that are produced by \textit{R. acidophilus}. Both species have very similar complements of LH complexes. Before measurements, $200$ $\mu$L of these suspensions were transferred to individual quartz cuvettes of 1 mm path length (CV1Q035AE, Thorlabs).

\noindent  \textbf{TWINS interferometer.} TWINS (manufactured by NIREOS s.r.l.) consists of two plates of a birefringent material, with thicknesses L$_a$ and L$_b$ respectively, and optical axes rotated by 90$\degree$. For an incident optical waveform with polarization rotated by 45$\degree$ with respect to the ordinary and extraordinary axes, the output waveform splits into two replicas with perpendicular polarizations and delay $\tau$ proportional to L$_b$-L$_a$. To continuously change the delay, one of the plates is cut into a pair of wedges, one of which is transversely translated. The two delayed replicas are then made to interfere by projecting them onto the same polarization with a polarizer. As it is a common path interferometer, TWINS guarantees exceptional delay stability and reproducibility.


\section*{References}
\bibliographystyle{naturemag}
\bibliography{bibliography}

\section*{Acknowledgements}
The authors acknowledge financial support from the Royal Academy of Engineering under the Chairs in Emerging Technology and Research Fellowships schemes and the U.K. Engineering and Physical Sciences Research Council (Grants No. EP/X035905/1, EP/Y029097/1, EP/Z533166/1).  G.C. and L.U. acknowledge support by the Progetti di ricerca di Rilevante Interesse Nazionale (PRIN) of the Italian Ministry of Research 2022HL9PRP Overcoming the Classical limits of ultRafast spEctroSCopy with ENtangleD phOtons (CRESCENDO). G.C. acknowledges funding from the European Union–NextGenerationEU under the National Quantum Science and Technology Institute (NQSTI) grant no. PE00000023-q-ANTHEM-CUP H43C22000870001.

\section*{Author Contributions Statement}
RA – methodology, investigation, formal analysis, visualization, writing, review and editing; LU- methodology, investigation, formal analysis, visualization, writing the original draft; AL - supervision, writing, review and editing; RJC – resources, writing, formal analysis, review and editing; GC - conceptualization, supervision, funding acquisition, writing the original draft; DF - conceptualization, supervision, funding acquisition, writing, review and editing.


\section*{Competing Interests Statement}
G.C. discloses financial association with the company NIREOS 
 (www.nireos.com), which manufactures the TWINS interferometer used in this paper.

\end{document}